\documentclass[12pt]{article}
\usepackage{graphicx}
\usepackage{amsmath}
\usepackage{longtable}

\begin{document}

\makeatletter
\renewcommand*{\@cite}[2]{{#2}}
\renewcommand*{\@biblabel}[1]{#1.\hfill}
\makeatother

\title{Spatial Variations of the Extinction Law in the Galactic Disk from Infrared Observations}
\author{G.~A.~Gontcharov\thanks{E-mail: georgegontcharov@yahoo.com}}

\maketitle

Pulkovo Astronomical Observatory, Russian Academy of Sciences, Pul\-kov\-skoe sh. 65, St. Petersburg, 196140 Russia


Infrared photometry in the $J$ (1.2 microns), $H$ (1.7 microns), $Ks$ (2.2 microns) bands from the 2MASS
catalogue and in the $W1$ (3.4 microns), $W2$ (4.6 microns), $W3$ (12 microns), $W4$ (22 microns) bands from the WISE
catalogue is used to reveal the spatial variations of the interstellar extinction law in the infrared near the
midplane of the Galaxy by the method of extrapolation of the extinction law applied to clump giants. The
variations of the coefficients $E_{(H-W1)}/E_{(H-Ks)}$, $E_{(H-W2)}/E_{(H-Ks)}$, $E_{(H-W3)}/E_{(H-Ks)}$
and $E_{(H-W4)}/E_{(H-Ks)}$
along the line of sight in $2^{\circ}\times 2^{\circ}$ squares of the sky centered at $b=0^{\circ}$
and $l=20^{\circ}, 30^{\circ}, 40^{\circ}, \ldots, 320^{\circ}, 330^{\circ}, 340^{\circ}$
as well as in several $4^{\circ}\times 4^{\circ}$ squares with $|b|=10^{\circ}$ are considered. The
results obtained here agree with those obtained by Zasowski et al. in 2009 using 2MASS and Spitzer-IRAC
photometry for the same longitudes and similar photometric bands, confirming their main result: in
the inner (relative to the Sun) Galactic disk, the fraction of fine dust increases with Galactocentric distance
(or the mean dust grain size decreases). However, in the outer Galactic disk that was not considered by
Zasowski et al., this trend is reversed: at the disk edge, the fraction of coarse dust is larger than that in
the solar neighborhood. This general Galactic trend seems to be explained by the influence of the spiral
pattern: its processes sort the dust by size and fragment it so that coarse and fine dust tend to accumulate,
respectively, at the outer and inner (relative to the Galactic center) edges of the spiral arms. As a result, fine
dust may exist only in the part of the Galactic disk far from both the Galactic center and the edge, while
coarse dust dominates at the Galactic center, at the disk edge, and outside the disk.

\newpage
\section*{INTRODUCTION}

The interstellar extinction law describes the dependence
of extinction $A_{\lambda}$ on wavelength $\lambda$. According
to present views, dust grains absorb radiation
with a wavelength smaller than their size. The
chemical composition, shape, and other properties of
dust grains also affect their interaction with radiation.
Therefore, determining the interstellar extinction law
and its spatial variations is important for refining
the properties of the dust and the entire interstellar
medium. These properties, in turn, tell about the
abundances of chemical elements heavier than helium
in the Universe, given that the bulk of the dust is
produced by stars. 

One of the most realistic models for the distribution
of interstellar dust grains in size and other
properties was proposed by Weingartner and Draine
(2001). The authors found that the dust grain size is
a key characteristic of the medium and that the versions
of the model envisaging the distribution of dust
grains in a wide range of sizes, at least from 0.001 to
6 microns, provide the best agreement with observations
in the solar neighborhood. For any three photometric
bands in this range, for example, $B$ (0.45 microns), $V$
(0.55 microns), and $K$ (2.2 microns), there exist dust grains
causing extinction only in the shortest-wavelength
band (in our example, these are dust grains smaller
than 0.5 microns), only in two bands (in our example,
smaller than 2 microns), and in all three bands (larger than
2.2 microns). In the latter case, if we restrict ourselves
to the data only in these three bands, then the extinction
will be perceived as wavelength-independent,
i.e., nonselective or ``gray''.

Near the Galactic plane in the solar neighborhood,
the extinction increases with decreasing wavelength,
because fine dust dominates (i.e., in our example, the
relationship between the extinctions in the $B$, $V$, and
$K$ bands is $A_{B}>A_{V}>A_{K}$). As a result, the radiation
passed through the dust reddens. In this case,
the ratio of the reddenings (color excesses) $E_{(V-K)}/E_{(B-V)}$
is a characteristic of the dust grain
size distribution in the range from 0.5 to 2.2 microns: a
lower value of this ratio means a larger extinction in $B$
than in $V$ and in $V$ than in $K$ and, hence, a steeper
extinction growth with decreasing wavelength and,
consequently, a larger fraction of fine dust grains
in their size distribution and a smaller mean grain
size. Conversely, a higher value of the ratio $E_{(V-K)}/E_{(B-V)}$
means a larger fraction of coarse dust
grains and a larger mean grain size. This provides a
basis for the classical method of determining an important
characteristic of the interstellar medium in the
solar neighborhood, the extinction coefficient $R_{V}\equiv A_{V}/E_{(B-V)}\approx1.1E_{(V-K)}/E_{(B-V)}$.
This is
the method of extrapolation of the extinction law or
the color-ratio method. The second name is justified
in that for stars with the same spectral energy distribution,
the ratio of the color excesses is equal to
the ratio of the colors, as was shown, for example, by
Gontcharov (2012a). Jonhson and Borgman (1963),
who were the first to apply this method, found not only
large deviations of $R_V$ from the mean value of 3.1 for
some stars but also a smaller (in amplitude) smooth
dependence of $R_V$ on Galactic longitude with its minimum
at $l\approx110^{\circ}$.

Ignoring the variations of the extinction law should
cause large systematic and random errors in calculating
the extinctions, distances, absolute magnitudes,
and other characteristics of stars. These errors as
a function of the error in $R_V$ were given by Reis
and Corradi (2008): in the case of $R_V$ variations of
$\pm1.5$ from the mean, the calculated distances and/or
magnitudes of stars are erroneous by 10\%. The systematic
variations of $R_V$ cause systematic errors of
the distances.

Skorzynski et al. (2003) specified several regions
of space in the solar neighborhood where significant
deviations of $R_V$ from the mean are obvious.

The $R_V$ variations within 600 pc of the Sun
were analyzed in more detail by Gontcharov (2012a).
Having applied the method of extrapolation of the
extinction law to multicolor broadband photometry
from the Tycho-2 catalogue (H\o g et al. 2000) in
the $B_T$ (0.435 microns) and $V_T$ (0.505 microns) bands and
the 2MASS (2-Micron All-Sky Survey) catalogue
(Skrutskie et al. 2006) in the $Ks$ (2.16 microns) band for
11 990 OB stars and 30 671 red giant branch (RGB)
KIII stars, we found consistent (for these classes
of stars) systematic spatial variations of $R_V$ in the
range from 2.2 to 4.4 and constructed a 3D map of
these variations with an accuracy $\sigma(R_V)=0.2$ and a
spatial resolution of 50 pc. The $R_V$ variations found
agree with the results by Skorzynski et al. (2003)
and are associated with Galactic structures: the
coefficient $R_V$ has a minimum at the center of the
Gould Belt, not far from the Sun; it reaches its
maximum at a distance of about 150 pc from the
center and then decreases to its minimum in the
outer part of the Gould Belt and other directions at
a distance of about 500 pc from the Sun, apparently
returning to the mean values far from the Sun. In
addition, a monotonic increase of $R_V$ toward the
Galactic center by 0.3 per kpc was found in a layer
$\sim200$ pc in thickness near the Galactic equator.
This result agrees with that obtained by Zasowski
et al. (2009) in the infrared (IR, $\lambda>1$ microns) for a much
larger part of the Galaxy. These systematic large scale
variations of the extinction law in the Galaxy,
if they are real, are the main result of the work by
Zasowski et al. (2009) and one of the most important
properties of the Galaxy found in the 21st century.
Therefore, their confirmation or refutation is needed,
which this paper is devoted to.

The coefficient $R_V$ characterizes the dust grain
size distribution in the range from 0.5 to 2.2 microns. This
distribution is often extrapolated to a wider range of
grain sizes. However, the same nonselective (gray)
extinction in all of the bands under consideration
arises in the case of a different dust grain size distribution,
for example, at a much larger fraction of
coarse dust. It can be revealed only by invoking even
longer-wavelength photometry. In fact, the present-day
attempts to estimate the nonselective extinction
are reduced to searching for the proofs that there are
no dust grains larger than some size in the Galaxy or,
at least, in a specific region of space.

The estimates of the size distribution of dust grains
larger than 2 microns are poor. The estimate by Weingartner
and Draine (2001) suggests an appreciable
fraction of dust grains up to 6 microns in size and a
corresponding mean abundance of chemical elements
heavier than helium that exceeds considerably the
value that usually follows from the observations of
stellar spectra. The requirements of a relatively high
abundance of carbon and silicon as the main elements
in the dust composition are particularly obvious.
The same authors found strong disagreement
of their model for any parameters with the observed
interstellar dust grain mass distribution in the Solar
system that was derived by Frisch et al. (1999)
from the detection of dust grain impacts with the
Ulysses and Galileo detectors: the theory gives a low
abundance of dust grains with a mass larger than
$2\times10^{-13}$ g, while observations show that dust grains
with masses from $10^{-12}$ to $2\times10^{-12}$ g (i.e., $\sim100$
billion atoms in one dust grain) make a major contribution
to the dust mass. Moreover, the Pioneer 10
and Pioneer 11 spacecraft detected a considerable
number of impacts by interstellar dust grains with
an even larger mass of $\sim2\cdot10^{-9}$ g. These dust
grains moved predominantly along the spacecraft --
Sun direction and could not be detected by Ulysses
and Galileo because of their unsuitable orientation
and orbital parameters (Kr\"uger et al. 2001). In all
these results, significantly different spatial distributions
and motions of interplanetary and interstellar
dust were found and taken into account. Consequently,
although the experiments were carried out
within the Solar system, the interstellar origin of these
dust grains is beyond doubt. Thus, contrary to the
previous views, coarse dust grains with masses of
$\sim10^{-12}\div10^{-9}$ g, i.e., with sizes of $\sim1-10$ microns, not
only can be widespread in the interstellar medium but
also can constitute the bulk of the dust.

Highly accurate multicolor IR photometry can
provide key data for estimating the abundance of
coarse dust in the interstellar medium, but it has
been obtained for millions of stars over the entire
sky only in recent years. The 2MASS catalogue
was produced in 2006 as a result of ground-based
observations in 1997--2001 and contains photometry
for more than 400 million stars over the entire sky
in the near-IR $J$ (1.2 microns), $H$ (1.7 microns), and $Ks$
(2.16 microns) bands. The most important results of
the GLIMPSE (Galactic Legacy Infrared Midplane
Survey Extraordinaire) survey, photometry for millions
of stars near the Galactic equator in four mid-
IR bands (traditionally called $\lambda_{3.6}$, $\lambda_{4.5}$, $\lambda_{5.8}$,
and $\lambda_{8.0}$ with the corresponding effective wavelengths of
3.545, 4.442, 5.675, and 7.76 microns), were obtained
by 2009 (Benjamin et al. 2003; Zasowski et al. 2009)
from the observations performed in 2003--2008 with
the IRAC camera of the Spitzer space telescope
(following Zasowski et al. (2009), we call these
data as Spitzer-IRAC ones). The WISE catalogue
(Wright et al. 2010) was produced in 2012 as a result
of the observations performed in 2010 with the Widefield
Infrared Survey Explorer telescope and contains
photometry for more than 500 million stars over the
entire sky, including almost all 2MASS stars in the
mid-IR $W1$ (3.4 microns), $W2$ (4.6 microns), $W3$ (12 microns),
and $W4$ (22 microns) bands.

The study by Zasowski et al. (2009) referring
almost exclusively to the inner (relative to the Sun)
Galaxy, where a combination of 2MASS and Spitzer-IRAC
photometry for clump giants was used, is an
example of successfully using highly accurate multicolor
IR photometry in the method of extrapolation
of the extinction law to detect large-scale spatial
variations of the extinction law and, consequently,
variations of the dust grain size distribution. Given
their width, the seven photometric bands used cover
the wavelength range from 1 to 10 microns. Clump giants
near the Galactic plane ($|b|<1.5^{\circ}$) in most of the first
and fourth Galactic quadrants farther than $10^{\circ}$ from
the Galactic center (a total of 290 square degrees in
the sky) in the range ofmagnitudes $11^{m}<J<15.5^{m}$
were selected on the color--magnitude $(J-Ks)$ -- $J$
diagram. These were selected at distances $1.8<r<15$ kpc from the Sun, excluding the Galactic
center region. The sky region under consideration
was divided into $2.5^{\circ}\times 2^{\circ}$ cells; from 820 to
60 000 selected stars fell into each of them. Color
indices with respect to the $H$ band like $(J-H)$,
$(H-Ks)$, etc. were used for each star. For all
stars in the cell, the linear dependences of $(J-H)$,
$(H-\lambda_{3.6})$, $(H-\lambda_{4.5})$, $(H-\lambda_{5.8})$,
and $(H-\lambda_{8.0})$ on $(H-Ks)$, which are equal to the ratios of the
color excesses and below are designated as $E_{(J-H)}/E_{(H-Ks)}$, $E_{(H-\lambda_{3.6})}/E_{(H-Ks)}$,
$E_{(H-\lambda_{4.5})}/E_{(H-Ks)}$, $E_{(H-\lambda_{5.8})}/E_{(H-Ks)}$, $E_{(H-\lambda_{8.0})}/E_{(H-Ks)}$,
were found by the least squares
method. These coefficients characterize
the IR extinction law and primarily the relative
abundance of supermicron-sized dust grains with
respect to the abundance of dust about 2 microns in size.
Since, in general, the dust grain size distribution
may not be a smooth monotonic function, all of the
mentioned coefficients can be independent of one
another and of the coefficient $R_V$ . Therefore, although
the systematic spatial variations of the IR extinction
law found by Zasowski et al. (2009) were recalculated
by the authors to the $R_V$ variations from 3.1 to 5.5,
this recalculation is illustrative in character until
the extinction law from the visual range to the IR
one is established for each region of space under
consideration.

Since the longitude in the Galactic region being
studied strongly correlates with the Galactocentric
distance, Zasowski et al. (2009) interpret the
longitude dependence of the extinction law as a
dependence on Galactocentric distance and, consequently,
a systematic decrease in dust grain size and
a decrease in nonselective extinction with increasing
Galactocentric distance. Thus, in their opinion,
supermicron- and submicron-sized dust grains dominate
in the central and outer regions of the Galaxy,
respectively. A correlation of the dust grain size with
the metallicity of stars, the chemical composition and
shape of dust grains, and other parameters that can
depend on Galactocentric distance is also possible.
Here, following Zasowski et al. (2009), it should
be noted that based on present views, we believe
the dust grain size to be the main factor affecting
the extinction law. However, if a different factor
(shape, chemical composition, etc.) will turn out to
be more important than the size in future, then all our
conclusions should be referred to this factor.

Zasowski et al. (2009) point out that even the
exclusion of well-known regions with anomalously
large $R_V$ from consideration does not remove the systematic
trend found. Thus, the variations of the extinction
law are actually inherent in a diffuse medium,
not in dense clouds, and the scale of these variations
is much larger than the well-known deviations of $R_V$
from 3.1 in small star-forming regions.

In this study, we extensively test the results by
Zasowski et al. (2009) using WISE data instead
of Spitzer-IRAC ones and by analyzing all Galactic
longitudes, not only the first and fourth quadrants. In
addition, we lay the groundwork for investigating the
dust properties outside the disk.

For the convenience of comparing the results, here
we consider color-excess ratios $E_{(H-\lambda)}/E_{(H-Ks)}$
similar to those in Zasowski et al. (2009), where
$\lambda$ are the four mid-IR photometric bands from the
WISE catalogue.

\section*{DATA REDUCTION}

The advantage of WISE over the Spitzer-IRAC
catalogue used by Zasowski et al. (2009) is the coverage
of the entire sky rather than its separate regions.
The main disadvantage of the WISE catalogue and
project is the unexpected loss of coolant earlier than
the planned time and, as a result, a short period
of observations and a decrease in the accuracy of
observations at the end of the period. However, the
decrease in accuracy is significant only in the $W3$
and particularly $W4$ bands. The accuracy of determining
the coefficients $E_{(H-W3)}/E_{(H-Ks)}$ and
$E_{(H-W4)}/E_{(H-Ks)}$ is approximately a factor
of 7 lower than that for $E_{(H-W1)}/E_{(H-Ks)}$ and
$E_{(H-W2)}/E_{(H-Ks)}$. Therefore, here greater
attention is paid to the results in the $W1$ and $W2$
bands.

For this study, we used only stars with an accuracy
of their photometry in the $J$, $H$, $Ks$, $W1$, and $W2$
bands better than $0.05^m$. This limitation rules out
not only faint stars but also the brightest ones. As
a result, clump giants within 700 pc of the Sun are
almost absent in our sample. However, despite the
photometric accuracy limitation, our sample is complete
or almost complete in a wide range of distances.
This range varies with longitude from $0.8-2.2$ to $0.8-6$ kpc.
All of the results presented below refer to this
range of distances. In the latter, our sample of clump
giants may not be complete only due to a slight loss
of stars that are members of nonsingle star systems,
for example, star pairs whose observed parameters do
not allow the clump giant to be identified.

The $W1$ (3.4 microns) and $W2$ (4.6 microns) fluxes can be
compared with those in the 3.6 and 4.5 microns bands from
the Spitzer-IRAC catalogue. The $W3$ band spans the
range from 7.5 to 17.5 microns and, therefore, the $W3$ flux
can be compared with that in the Spitzer-IRAC 8 microns
band.

Thus,we have photometry formany stars over the
entire sky in seven bands: three from 2MASS and
four from WISE spanning the range from 1 to about
27 microns, given the band width.

Analysis shows that under different conditions,
different combinations of photometric bands are the
best characteristics of the dust grain size distribution
and nonselective extinction estimates. The method of
extrapolation of the extinction law requires that, first,
a representative, desirably complete sample of stars
with the same unreddened spectral energy distribution
be available in the cell of space under consideration,
second, the reddening and extinction gradient
within this cell be larger than the photometric errors,
and, third, at least one of the colors should redden
appreciably within the cell to compare the reddening
and nonselective extinction, i.e., the reddening
itself and extinction should be sufficiently large. The
contradiction between these conditions stems from
the fact that (1) the reddening mixes stars of different
classes when any photometric characteristics
are considered, (2) the extinction hides stars, and
(3) the reddening and extinction are low when only
IR photometry is used. Therefore, in regions with
particularly large reddening and extinction, for example,
near the Galactic plane in the first and fourth
Galactic quadrants, it is most important to achieve
sample completeness. Consequently, the pairs of
colors composed of the longest-wavelengths IR magnitudes,
with one pair still showing an appreciable
reddening, are most suitable, i.e., the choice by Zasowski
et al. (2009), $E_{(H-\lambda)}/E_{(H-Ks)}$, where $\lambda$
from the mid-IR range, is justified.

In regions with moderate extinction, for example,
in the Galactic disk in the second and third quadrants,
an accurate reddening estimate is more important.
Therefore, it is more advantageous to use a color that
reddens more strongly, i.e., $(J-H)$ is more suitable
than $(H-Ks)$. Here, for comparison with the results
by Zasowski et al. (2009), we used the $(H-Ks)$
color, although all our calculations were also performed
using $(J-H)$, which yielded qualitatively the
same results.

Since the IR reddening is negligible far from the
Galactic plane, one of the colors used should contain
visual bands, i.e., for example, the classical $E_{(V-Ks)}/E_{(B-V)}$ is suitable.

A comprehensive analysis of the extinction law
based on WISE and 2MASS data requires processing
more than 1 Tb of information and can be given
in several publications. Here, we perform a preliminary
analysis -- we consider $2^{\circ}\times 2^{\circ}$ squares of the sky
centered at $b=0^{\circ}$ and $l=20^{\circ}, 30^{\circ}, 40^{\circ}, \ldots, 320^{\circ}, 330^{\circ}, 340^{\circ}$,
i.e., along the entire Galactic equator, except for the
directions near the Galactic center. In addition, we
analyze several $4^{\circ}\times 4^{\circ}$ squares with $|b|=10^{\circ}$.

Just as was done by Zasowski et al. (2009), here
we selected clump red giants, i.e., evolved stars with
nuclear reactions in the helium core. Their evolutionary
status and characteristics were considered
in detail by Gontcharov (2008) when analyzing a
sample of 97 348 such stars from the Hipparcos
(ESA, 1997; van Leeuwen 2007) and Tycho-2 catalogues
mostly within 1 kpc of the Sun. In particular,
their empirical mean absolute magnitudes derived by
Gontcharov (2008) using Hipparcos parallaxes agree,
within $0.05^m$, with other estimates, for example,
from Groenewegen (2008), and with the theoretical
estimates from the Padova database of evolutionary
tracks (http://stev.oapd.inaf.it/cmd; Bertelli
et al. 2008; Marigo et al. 2008). These theoretical
estimates used below for a mixture of clump giants
with metallicities $Z$ from 0.004 to 0.020 and masses
from 0.9 to 3 $M_{\odot}$ at a mean metallicity $Z\approx0.014$
(i.e., $FeH\approx-0.13$, slightly lower than the solar one)
and a mass of 1.4 solar masses give $\overline{M_{J}}=-0.93^{m}$,
$\overline{M_{H}}=-1.44^{m}$, $\overline{M_{Ks}}=-1.52^{m}$,
$\overline{M_{W1}}=-1.55^{m}$, $\overline{M_{W2}}=-1.49^{m}$, $\overline{M_{W3}}=-1.58^{m}$, $\overline{M_{W4}}=-1.59^{m}$.
Variations in the metallicity and mass of clump giants
in a wide, but reasonable range lead to variations in
the above means within $\pm0.1^{m}$.

In agreement with the Padova database, Gontcharov
(2008) showed that $M_{Ks}$ is almost independent
of the colors and that the scatter of individual values
of $M_{Ks}$ about the mean is $0.3^m$. This allows the
distance $r$ for each selected star to be calculated with
a 15\% accuracy:
\begin{equation}
\label{rph}
r=10^{(Ks+1.52+5-A_{Ks})/5},
\end{equation}
where the extinction
\begin{equation}
\label{ajk}
A_{Ks}=0.6E_{(J-Ks)}=0.6((J-Ks)-(J-Ks)_0)=0.6((J-Ks)-0.62),
\end{equation}
given that $\overline{(J-Ks)_0}=0.62^m$ for unreddened stars.
Here, we adopted the coefficients
\begin{equation}
\label{akejk}
A_{Ks}/E_{(J-Ks)}=0.6, A_{Ks}/A_{V}=0.1
\end{equation}
as the means from many extinction laws (Cardelli
et al. 1989; Indebetouw et al. 2005; Marshall et al.
2006; Nishiyama et al. 2009; Draine 2003; and
references therein). The uncertainty in these coefficients
reaches 10\%, but this is an acceptable value,
because everywhere $E_{(J-Ks)}<1^{m}$. Therefore, the
uncertainty $\sigma(A_{Ks})<0.06^{m}$, and this gives a relative
error in $r$ of no more than 5\%, i.e., much less than the
error in $r$ due to the scatter of $M_{Ks}$.

On the Hertzsprung–Russell diagram, the clump
giants have
\begin{equation}
\label{mjkm}
0.55^{m}<(J-Ks)_0<0.9^{m}
\end{equation}
adjoining and partially intersecting with the RGB
and AGB stars ($(J-Ks)_0>0.7^{m}$), bright giants
and supergiants ($M_{Ks}<-2.5^{m}$), subgiants ($M_{Ks}>-0.5^{m}$), and red dwarfs ($M_{Ks}>3.5^{m}$).
Gontcharov
(2008) showed that clump giants dominate among
stars of the same color and absolute magnitude
(forming precisely a clump). Gontcharov (2009a)
confirmed that the RGB and AGB stars, bright
giants, supergiants, and subgiants give small admixtures
in the sample of clump giants when they are
selected by color. In addition, it was shown that using
reduced proper motions (see below) allows the clump
giants to be separated from the red dwarfs of the same
color.

Just as was done by Zasowski et al. (2009),
here we selected the clump giants as all stars in
the region of enhanced star density on the color–
magnitude diagram, but we used $(J-Ks)$ -- $Ks$
instead of $(J-Ks)$ -- $J$ in Zasowski et al. (2009).
This selection method was first applied by Lopez-Corredoira et al. (2002) and was developed by Drimmel
et al. (2003), Indebetouw et al. (2005), and
Marshall (2006). Gontcharov (2010) showed that
to properly identify the clump giants on the $(J-Ks)$ -- $Ks$ diagram, the reddening of stars should be
estimated and the sample should be cleaned from the
admixture of red dwarfs with $(J-Ks)_0<0.9^m$.

Just as was done by Zasowski et al. (2009), we
assume that the clump giants have $0.55^{m}<(J-Ks)_0<0.9^{m}$. The reddening smears and shifts this
range. Since $b\approx0$, we assume that
\begin{equation}
\label{akr}
A_{Ks}=k\cdot r,
\end{equation}
where $k$ is the coefficient determined for each sky region
under consideration so that the maximum number
of stars fall into a fixed selection region on the
$(J-Ks)$ -- $Ks$ diagram. Thus, we select precisely the
clump of giants, i.e., the region of their enhanced density
on the diagram. Since in this approach $A_{Ks}$ and
$r$ in Eqs. (1) and (5) depend on each other, they are
refined by iterations. The coefficients $k$ found agree
well with the known extinction variations with longitude
pointed out, for example, by Gontcharov (2009b,
2012b).

An example of the $(J-Ks)$ -- $Ks$ diagram for the
$2^{\circ}\times 2^{\circ}$ square of the sky centered at $b=0^{\circ}$ and
$l=100^{\circ}$ is given in Fig. 1. The left cloud of stars
consists of O-F main-sequence stars, the central
cloud consists of clump giants, with the fraction of
dwarfs being large among the faintest stars (the stars
in the region of the diagram where the red dwarfs
dominate are marked by the gray symbols), and the
stars in the right part of the diagram are RGB ones.
The selection boundaries derived for this square of the
sky are indicated by two curves that are ultimately
defined by Eqs. (1)--(5); more specifically, for the
brightest stars with negligible extinction, they are
defined by Eq. (4), while the shift of the curves along
the horizontal axis with increasing $Ks$ depends on
the coefficient $k$ and changes from one square of the
sky to another. The white straight line at the bottom
of the figure cuts off the diagram region where the
clump giants are inseparable from the RGB stars and
dwarfs, the sample of clump giants is incomplete, or
the effects distorting the result are large.

The efficient of identifying the clump giants is
explained by the following reasoning. We select the
clump giants with $r<5$ kpc. At $M_{Ks}=-1.52^{m}$,
we then have either (1) $A_{Ks}>0.2^{m}$ and $(J-Ks)>0.9^{m}$ or
(2) $A_{Ks}<0.2^{m}$ and $Ks<12^{m}$. The dwarfs
that can become an admixture because of their proximity
to the Sun have $A_{Ks}\approx0$ and $0.55^{m}<(J-Ks)<0.9^{m}$. Condition (1) is then reflected at the
bottom of Fig. 1 -- the reddening of giants does not allow
them to be mixed with dwarfs because of different
colors, while condition (2) is reflected in the middle
part of the figure -- the dwarfs with $M_{Ks}\approx4.5^{m}$ at
$Ks<12^{m}$ are within $r<300$ pc, while the number
of dwarfs in this space is smaller than the number of
clump giants within 5 kpc of the Sun approximately
by one or two orders of magnitude.

The red dwarfs that fell into the sample were
excluded as stars with large reduced proper motions
$M'_{Ks}=Ks-A_{Ks}+5+5\cdot~\lg(\mu)$, where $\mu=(\mu_{\alpha}\cos\delta^2+\mu_{\delta}^2)^{1/2}$
is the total proper motion in arcseconds
taken from the PPMXL catalogue (Roeser
et al. 2010), where it is available for most of the stars
considered. At $b=0^{\circ}$, the dwarfs revealed in this
way account for less than 1\% of the preselected stars.
However, using $M'_{Ks}$
turned out to be very important
at $|b|=10^{\circ}$ (and, undoubtedly, extremely important
at high latitudes): here, the admixture of dwarfs is
about 10\% and can apparently be revealed only by
using the reduced proper motions.

Since the stars were selected using $(J-Ks)$,
although the spatial variations of $E_{(J-H)}/E_{(H-Ks)}$ were analyzed by Zasowski et al. (2009), this
analysis is distorted by strong selection effects and is
not considered here.

In each square of the sky under consideration,
we performed a moving calculation of the coefficients
$E_{(H-W1)}/E_{(H-Ks)}$, $E_{(H-W2)}/E_{(H-Ks)}$, $E_{(H-W3)}/E_{(H-Ks)}$, $E_{(H-W4)}/E_{(H-Ks)}$
as a function of $r$ with an averaging window
dependent on the star density in a given direction
and varied in the range from 100 (toward the Galactic
anticenter) to 300 (toward the Galactic center)
stars. The stars are arranged by $r$ and the mean $\overline{r}$,
along with the sought-for coefficients, is calculated
for $100-300$ stars with minimum $r$. Then, we exclude
the star with minimum $r$ from the set of stars under
consideration, introduce the previously unused star
with minimum $r$ instead of it, and repeat the calculations
of $\overline{r}$ and the sought-for parameters. As a result,
we obtain more than a thousand solutions including $\overline{r}$
with the corresponding set of sought-for coefficients
for each square of the sky under consideration.

As has been pointed out above, the relative accuracy
of individual $r$ determined by the scatter of
individual $M_{Ks}$ about the mean is 15\%. The relative
accuracy of $\overline{r}$ is then $1-1.5\%$, i.e., given that the
number of stars increases with distance in the solid
angles under consideration, $15-60$ pc. Therefore, it
is hoped that the variations of the coefficients under
consideration in intervals larger than 60 pc are real.

\section*{RESULTS}

The spatial variations of the coefficients $E_{(H-W1)}/E_{(H-Ks)}$, $E_{(H-W2)}/E_{(H-Ks)}$, $E_{(H-W3)}/E_{(H-Ks)}$
revealed here agree with
those found by Zasowski et al. (2009) for the analogous
coefficients $E_{(H-\lambda_{3.6})}/E_{(H-Ks)}$, $E_{(H-\lambda_{4.5})}/E_{(H-Ks)}$,
$E_{(H-\lambda_{8.0})}/E_{(H-Ks)}$,
although Zasowski et al. (2009) provide only the
mean values of the coefficients for each longitude
and, in addition, the ranges of distances considered by
Zasowski et al. (2009) and in our study are different.

As an example, Fig. 2 shows the variations of
$E_{(H-W2)}/E_{(H-Ks)}$ with heliocentric distance
for longitudes of $40^{\circ}$, $50^{\circ}$, $60^{\circ}$, $260^{\circ}$, $280^{\circ}$,
$290^{\circ}$, $300^{\circ}$, $310^{\circ}$, $320^{\circ}$ (black curves with gray error bands)
in comparison with the mean coefficient $E_{(H-\lambda_{4.5})}/E_{(H-Ks)}$ from Zasowski et al. (2009) (horizontal
straight lines; the accuracy of the mean
result by Zasowski et al. (2009) is approximately
the thickness of these straight lines). Comparison
for longitudes of $20^{\circ}$, $30^{\circ}$, $90^{\circ}$, $270^{\circ}$, $330^{\circ}$, $340^{\circ}$
is given in other figures. We see that some of the
variations in coefficient are significant and seem to
correspond to large Galactic structures, most likely
spiral arms. We also see that, despite the variations
in an interval of tens and hundreds of pc, there are
no systematic $E_{(H-W2)}/E_{(H-Ks)}$ trends in
an interval of several kpc at these longitudes in the
$W2$ band.

In Fig. 3, the coefficients $E_{(H-W1)}/E_{(H-Ks)}$, $E_{(H-W2)}/E_{(H-Ks)}$, $E_{(H-W3)}/E_{(H-Ks)}$,
$E_{(H-W4)}/E_{(H-Ks)}$ are plotted
against the heliocentric distance for longitudes of
$200^{\circ}$ (negative $r$) and $20^{\circ}$ (positive $r$) (Figs. 3a--3d),
$160^{\circ}$ (negative $r$) and $340^{\circ}$ (positive $r$) (Figs. 3e--3h))
(black curves with gray error bands) in comparison
with the analogous results from Zasowski
et al. (2009) (horizontal straight lines; the accuracy
of the mean is approximately the thickness of these
straight lines). Figure 4 shows analogous results for
longitudes of $210^{\circ}$ (negative $r$) and $30^{\circ}$ (positive $r$)
(Figs. 4a–4d), $150^{\circ}$ (negative $r$) and $330^{\circ}$ (positive
$r$) (Figs. 4e--4h). These longitudes were selected
among all of those considered as being closest to the
direction to the Galactic center ($20^{\circ}$, $30^{\circ}$, $330^{\circ}$ and
$340^{\circ}$) and oppositely directed with respect to them
($200^{\circ}$, $210^{\circ}$, $150^{\circ}$ and $160^{\circ}$, respectively).

The minima in the curves shown in the figure correspond
to a larger fraction of dust grains with sizes of
about 3, 4, 10, and 20 microns than that of dust grains with
a size of about 2 microns for $E_{(H-W1)}/E_{(H-Ks)}$, $E_{(H-W2)}/E_{(H-Ks)}$, $E_{(H-W3)}/E_{(H-Ks)}$,
$E_{(H-W4)}/E_{(H-Ks)}$, respectively, or other
dust properties causing a higher extinction of radiation
with a longer wavelength. In contrast, the maxima
in the curves correspond to a larger amount of fine
dust if precisely the dust grain size is a characteristic
determining the extinction.

In Figs. 3 and 4, just as in Fig. 2, we see agreement
of our results with those from Zasowski et al. (2009).
In addition, there is a clear correlation between the
results in the $W1$ and $W2$ bands and their slight
deviation from those in the $W4$ and particularly $W3$
bands.

For $W4$, this discrepancy is due to a large deviation
of the wavelength from the $W1$ and $W2$ bands.
This leads us to conclude that the fraction of dust
grains about 20 microns in size represented by the coefficient
$E_{(H-W4)}/E_{(H-Ks)}$ and their spatial distribution
correlate weakly with those for dust grains
about 3 and 4 microns in size.

The results for the $W3$ band differ from the remaining
ones, because there is a well-known broad
absorption line with $\lambda=9.7$ microns in this band produced
by amorphous silicates contained in the dust.
Indebetouw et al. (2005) and Zasowski et al. (2009)
discussed this effect for the Spitzer-IRAC $\lambda=8$ microns
band and showed an insignificant manifestation of
this effect for shorter-wavelength bands. The absence
of a correlation between the results for $W3$ and $W4$ in
our study shows that this effect apparently does not
manifest itself at $\lambda>20$ microns either. In the $W3$ band in
Figs. 3 and 4, we see a systematic trend of the curves
absent for the $W1$, $W2$, and $W4$ bands. This trend is
consistent for all four longitudes under consideration
and exhibits a minimum of $E_{(H-W3)}/E_{(H-Ks)}$
within the nearest kiloparsec. Since the data acquisition
and reduction procedure in all bands is the same,
this trend seems to be not an artifact but a manifestation
of a peculiar distribution and/or properties
of the dust responsible for the extinction in the $W3$
band. Given the probable connection of this extinction
with amorphous silicates, it can be hypothesized
that, in contrast to the remaining coefficients, the
spatial variations of $E_{(H-W3)}/E_{(H-Ks)}$ seen in
the figures are caused not by variations of the dust
grain size distribution but by variations of the dust
mean chemical composition.

For all of the longitudes under consideration, except
perhaps $160^{\circ}$ and $340^{\circ}$ (Fig. 3h), the same systematic
trend of the curves approximately illustrated
by the dashed curves can be seen in the $W1$, $W2$, and
$W4$ bands: as one recedes from the Galactic center,
the curves reach a global maximum near $r\approx1-2$ kpc
(slightly closer to the Sun in $W4$ than in $W1$ and $W2$)
and in this part they agree with the results from Zasowski
et al. (2009), but still farther from the Galactic
center, despite the local maxima and minima,
the coefficients under consideration tend to decrease
(i.e., the gray extinction in the solar neighborhood is
minimal). If the spatial variations of the coefficients
under consideration are caused by variations of the
dust grain size distribution, then this systematic trend
shows a decrease in the fraction of coarse dust with
increasing Galactocentric distance in the inner (relative
to the Sun) Galaxy (the same conclusion for the
same region was reached by Zasowski et al. (2009))
and a decrease in this fraction in the outer Galaxy.

The global extrema and trends are less prominent
in Fig. 5, where results similar to those in Fig. 3 are
shown, but for longitudes of $270^{\circ}$ (negative $r$) and
$90^{\circ}$ (positive $r$): the results for $W3$ have a minimum
within the nearest kiloparsec here as well, while for
a longitude of $90^{\circ}$, there is a noticeable decrease in
$E_{(H-W1)}/E_{(H-Ks)}$ and $E_{(H-W2)}/E_{(H-Ks)}$ with increasing heliocentric distance, especially
farther than 5 kpc, although both effects are less
pronounced than in Figs. 3 and 4. The results for
adjacent longitudes in the $W2$ band presented in
Figs. 2c--2f show no clear extrema and trends, although
the trends will possibly manifest themselves
farther than 5 kpc from the Sun for all these longitudes,
just as for $l=90^{\circ}$. Our results for $l=90^{\circ}$ in
the $W1$ and $W2$ bands, on average, differ markedly
from those from Zasowski et al. (2009). However,
this is the only longitude with such a difference and,
in addition, the result from Zasowski et al. (2009) for
$l=90^{\circ}$ differs from their results for all the remaining
longitudes.

Significant rises and falls of all coefficients are
clearly seen at a longitude of $270^{\circ}$, in which the dustrich
regions of the local spiral arm seem to manifest
themselves. In any case, the extinction begins to
sharply increase at this longitude at a distance of
more than 1 kpc. This can be seen when analyzing
the corresponding $(J-Ks)$ -- $Ks$ diagram: $A_{Ks}\approx0.7^m$ already at a heliocentric distance of 2.5 kpc,
i.e., $A_{V}\approx7^m$, according to Eqs. (3). High extinction
begins to hide clump giants, and the sample becomes
incomplete. Therefore, no reliable results have been
obtained farther than 2.5 kpc from the Sun for a
longitude of $270^{\circ}$.

Thus, the global extrema of the coefficients under
consideration manifest themselves at certain Galactocentric
distances, within $\pm3$ kpc of the Sun's
Galactocentric distance. The Galactic disk at these
Galactocentric distances differs from the remaining
space of the Galaxy primarily by the great role of the
spiral arms that are absent at the Galactic center, at
the disk edge, and outside the disk.

Figure 6 shows the same as in Figs. 3--5 but for
the squares of the sky centered at
$l=180^{\circ}$, $b=0^{\circ}$ (black curves, negative $r$),
$l=180^{\circ}$, $b=+10^{\circ}$ (gray solid curves, negative $r$),
$l=180^{\circ}$, $b=-10^{\circ}$ (gray dashed curves, negative $r$),
$l=340^{\circ}$, $b=0^{\circ}$ (black curves, positive $r$),
$l=340^{\circ}$, $b=+10^{\circ}$ (gray solid curves, positive $r$),
$l=340^{\circ}$, $b=-10^{\circ}$ (gray dashed curves, positive $r$).
We see that the variations of
$E_{(H-W1)}/E_{(H-Ks)}$ and $E_{(H-W2)}/E_{(H-Ks)}$ change qualitatively even at a small distance
from the Galactic plane: toward both the Galactic
center and anticenter, both above and below the
Galactic plane, these coefficients are systematically
much smaller than those in the plane itself, although
the curves approach each other toward the Galactic
anticenter. If precisely the dust grain size is a characteristic
determining the extinction law, then we see
that the fraction of fine dust in the Galactic plane is
maximal (or the mean dust grain size is minimal), but
the fraction of coarse dust (or the mean grain size)
increases with distance from the plane. Where the
finest dust is observed in the plane, the fraction of
coarse dust outside the plane is particularly large: the
black and gray curves anticorrelate with each other in
many places.

In the inner (relative to the Sun) Galaxy (positive
$r$ in the figures), the difference between the dust grain
sizes in and outside the Galactic plane also manifests
itself for $E_{(H-W3)}/E_{(H-Ks)}$ and $E_{(H-W4)}/E_{(H-Ks)}$.

Sharp drops of the curves at $r\approx-1$ kpc and $r\approx-2.8$ kpc are clearly seen in Figs. 3a, 3b, 3d--3f, 3h,
4a, 4b, 4e, 4f, and 4h. These apparently show the
positions and influence of spiral arms, respectively,
the Perseus arm and the more distant outer arm. Different
arms seem to have also manifested themselves
in the same way on other graphs, but, in many cases,
an improper perspective or errors in the distances of
stars smooth out their manifestation. We see that, in
most cases, at a sharp drop of the curve, its minimum
is farther from the Galactic center than its maximum.
Thus, coarser dust dominates in the outer (relative to
the Galactic center) part of the arm, while fine dust
dominates in its inner part. This is an expected result
for a twisting spiral pattern with a leading arm edge:
an increase in the density of the interstellar medium
and other processes at the outer edge lead to intense
sticking of dust grains, while supernova explosions
and other processes deep in the arm destroy the dust
grains. Thus, the disk at Galactocentric distances
of $4-12$ kpc (if the Sun is 8 kpc away from the
Galactic center), where the spiral pattern exists, is a
Galactic region (possibly the only one) where the dust
is sorted by size and is fragmented significantly. This
``dust sorting and fragmentation factory'' apparently
does not work where there is no spiral pattern, i.e.,
near the Galactic center (as was shown by Zasowski
et al. 2009), at the disk edge (farther than 5 kpc from
the Sun toward the Galactic anticenter, according to
Figs. 6a and 6b), and outside the disk (gray curves in
Figs. 6a and 6b). In all these regions outside the spiral
pattern, the dust grain size distribution must then be
the same, constant, and noticeably different from that
in the solar neighborhood apparently in favor of coarse
dust grains. This will be verified in our subsequent
study of the extinction law and related dust properties
\emph{outside the Galactic disk} using 2MASS and WISE
data.

\section*{CONCLUSIONS}

This study showed the possibility of using the new
WISE catalogue, along with the 2MASS catalogue,
to analyze the interstellar dust properties and the corresponding
variations of the IR extinction law based
on multicolor IR photometry in the range from 1
to 27 microns for millions of stars over the entire sky.
Owing to the high accuracy of the photometric data
and the very large number of stars used, the classical
method of extrapolation of the extinction law
applied to clump giants is operable. The coefficients
$E_{(H-W1)}/E_{(H-Ks)}$, $E_{(H-W2)}/E_{(H-Ks)}$, $E_{(H-W3)}/E_{(H-Ks)}$ and $E_{(H-W4)}/E_{(H-Ks)}$
obtained in this method are good characteristics
of the extinction law and dust properties near the
Galactic plane, although the analogous coefficients
using $E_{(J-H)}$ instead of $E_{(H-Ks)}$ are also informative
in the second and third quadrants.

It turned out that the spatial variations of these coefficients
for the $W1$ and $W2$ bands correlate with one
another and usually differ slightly from those for the
$W4$ band and markedly for the W3 band. The results
for the $W3$ band differing from the remaining ones
seem to be affected by the absorption by amorphous
silicates at a wavelength of 9.7 microns and, in contrast to
the remaining bands, seem to reflect the variations of
the mean chemical composition rather than the dust
grain size to a greater extent.

Our results agree with those obtained by Zasowski
et al. (2009) using 2MASS and Spitzer-IRAC photometry
for the same longitudes and similar photometric
bands, confirming the main result by Zasowski
et al. (2009): in the inner (relative to the Sun)
Galactic disk, the fraction of fine dust increases with
Galactocentric distance (or the mean dust grain size
decreases). However, in the outer Galactic disk that
was not considered by Zasowski et al. (2009), this
trend is reversed: at the disk edge, the fraction of
coarse dust is larger than that in the solar neighborhood.
This general Galactic trend seems to be
explained by the influence of the spiral pattern: its
processes sort the dust by size and fragment it so that
coarse and fine dust tend to accumulate, respectively,
at the inner and outer (relative to the Galactic center)
edges of the spiral arms. As a result, fine dust may
exist only in the part of the Galactic disk far from both
the Galactic center and the edge, while the fraction
of fine dust at the Galactic center, the disk edge, and
outside the disk is minimal.

\section*{ACKNOWLEDGMENTS}

In this study, we used results from the Two
Micron AllSky Survey (2MASS), Wide-field Infrared
Survey Explorer (WISE), and PPMXL projects as
well as resources from the Strasbourg Astronomical
Data Center (Centre de Donn\'ees astronomiques de
Strasbourg). This study was supported by Program
P21 of the Presidium of the Russian Academy of Sciences
and the ``Scientific and Scientific--Pedagogical
Personnel of Innovational Russia'' Federal Goal-Oriented Program, XXXVII queue—Activity 1.2.1.

\newpage

\begin{figure}
\includegraphics{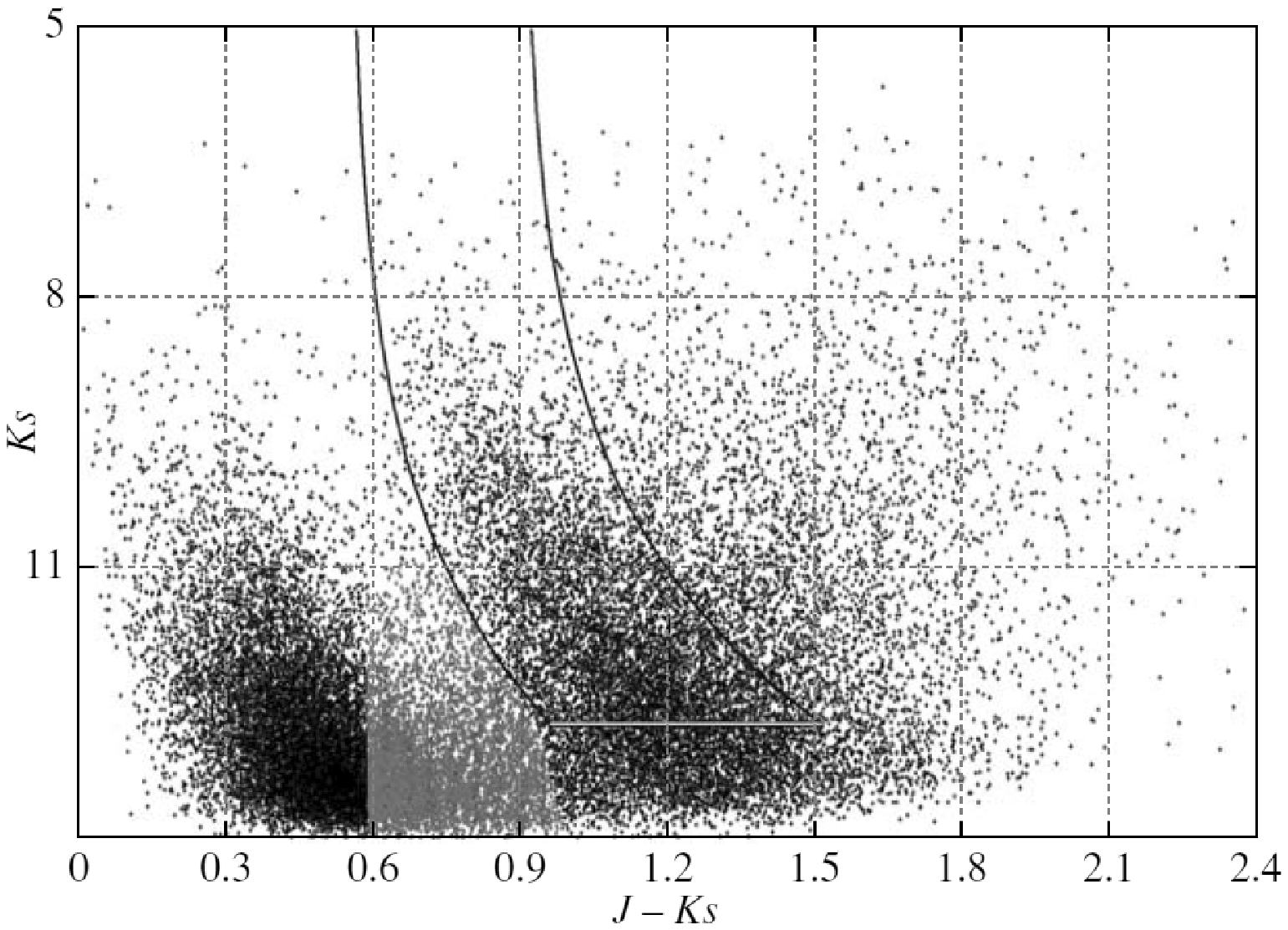}
\caption{Distribution of the stars under consideration on the $(J-Ks)$ -- $Ks$ diagram in the $2^{\circ}\times 2^{\circ}$
sky region centered at
$l=100^{\circ}$, $b=0^{\circ}$. All stars in the region bounded by the curves and horizontal straight line were selected. The gray symbols
mark the stars in the diagram region where red dwarfs of the same color as that of the selected clump giants dominate.
}
\label{jkk}
\end{figure}

\begin{figure}
\includegraphics{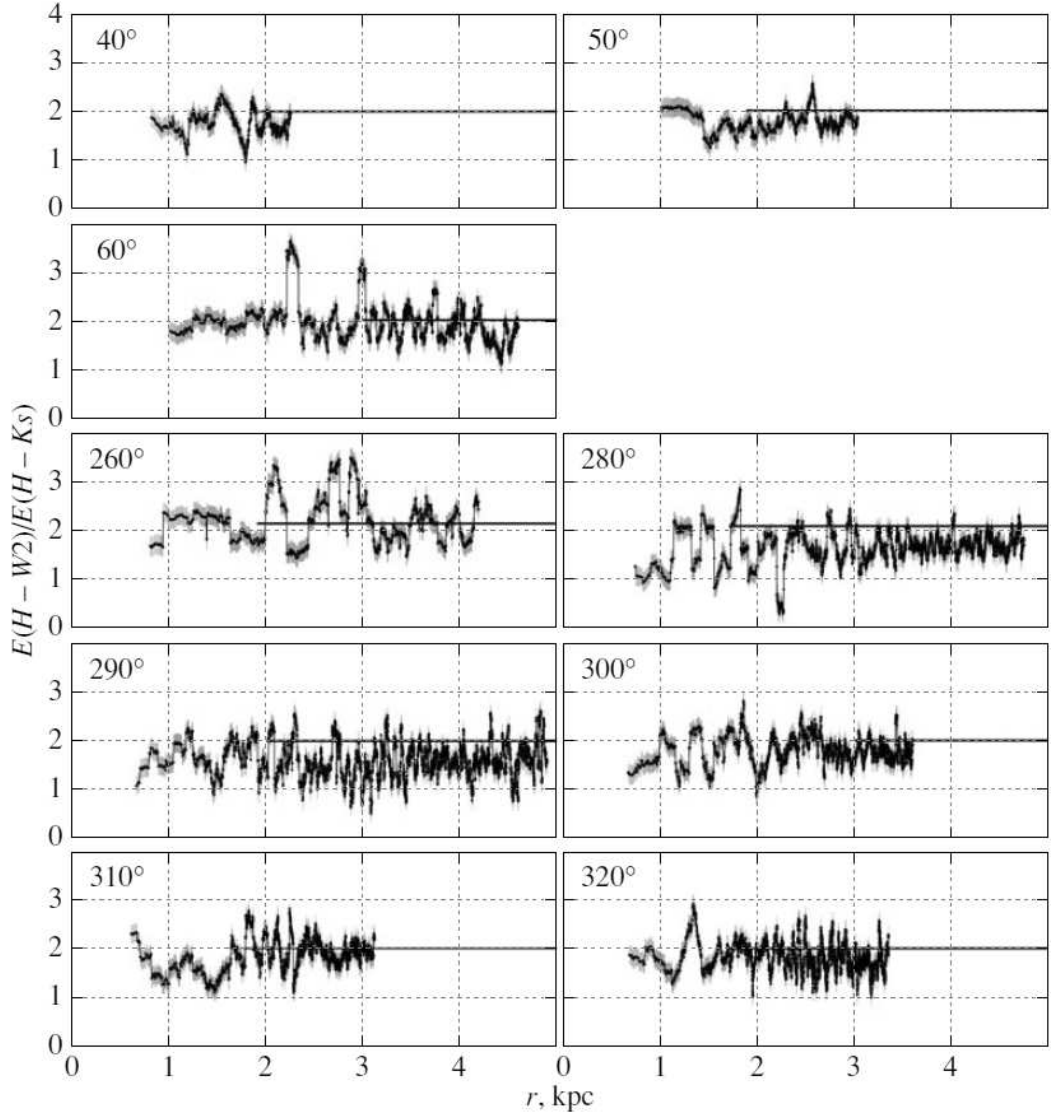}
\caption{$E_{(H-W2)}/E_{(H-Ks)}$ versus heliocentric distance for longitudes of $40^{\circ}$,
$50^{\circ}$, $60^{\circ}$, $260^{\circ}$, $280^{\circ}$, $290^{\circ}$,
$300^{\circ}$, $310^{\circ}$, $320^{\circ}$ (black curves with gray error bands)
in comparison with the analogous results from Zasowski et al. (2009) (horizontal
straight lines).
}
\label{wisespitzer}
\end{figure}

\begin{figure}
\includegraphics{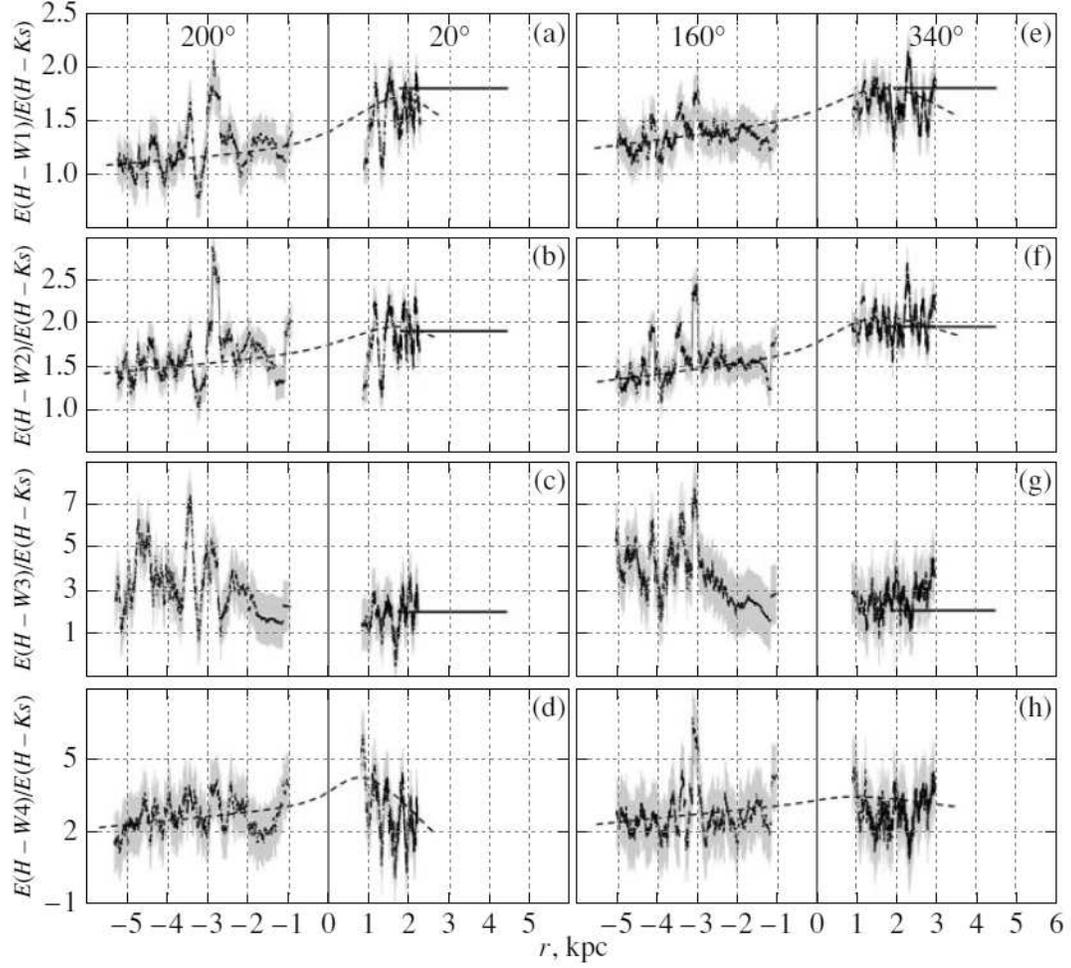}
\caption{$E_{(H-W1)}/E_{(H-Ks)}$, $E_{(H-W2)}/E_{(H-Ks)}$, $E_{(H-W3)}/E_{(H-Ks)}$, $E_{(H-W4)}/E_{(H-Ks)}$
versus heliocentric distance for longitudes of $200^{\circ}$ (negative $r$) and
$20^{\circ}$ (positive $r$) (a--d), $160^{\circ}$ (negative $r$) and $340^{\circ}$
(positive $r$) (e--h) (black curves with gray error bands).
The horizontal straight lines indicate the analogous results from
Zasowski et al. (2009). The dashed lines indicate the systematic variations of the coefficients.
}
\label{rl20}
\end{figure}

\begin{figure}
\includegraphics{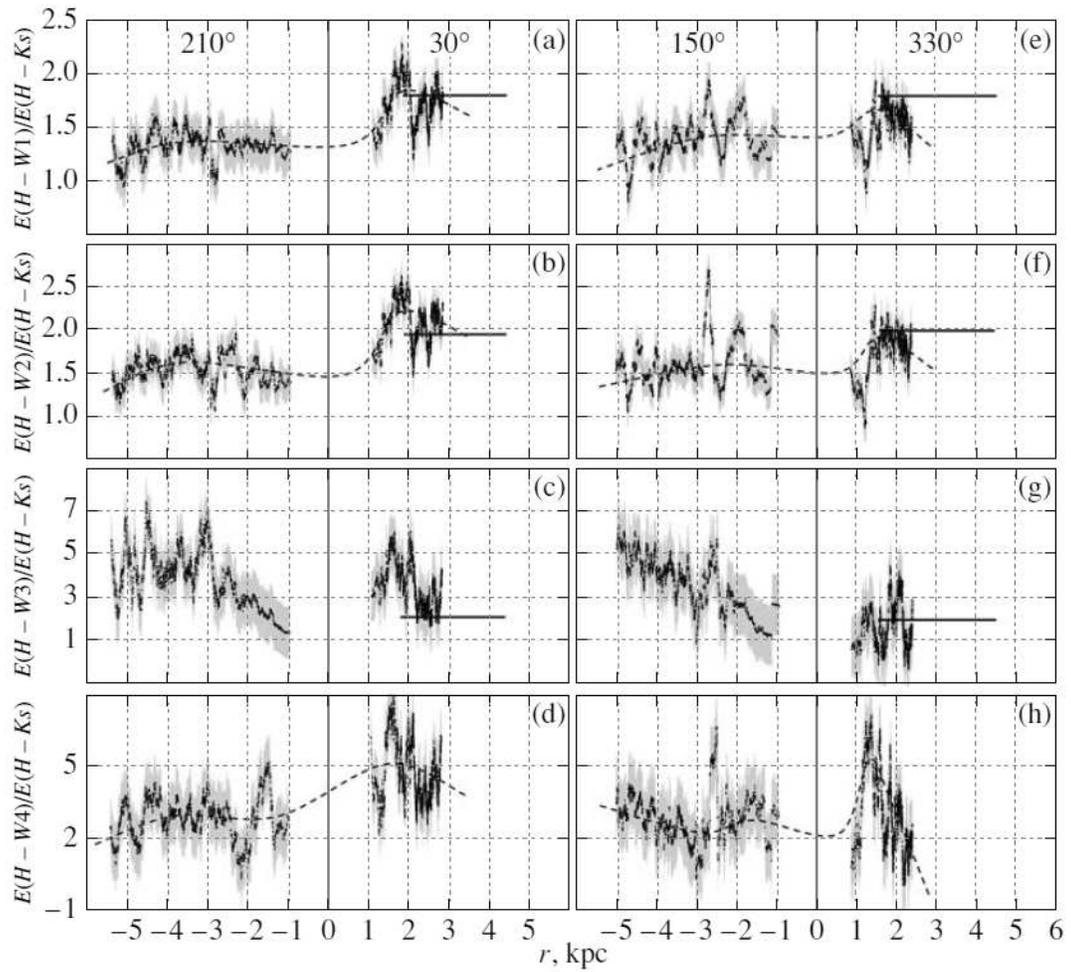}
\caption{Same as Fig. 3 for longitudes of $210^{\circ}$ (negative $r$) and $30^{\circ}$ (positive $r$)
(a--d), $150^{\circ}$ (negative r) and $330^{\circ}$ (positive r)
(e--h).
}
\label{rl30}
\end{figure}

\begin{figure}
\includegraphics{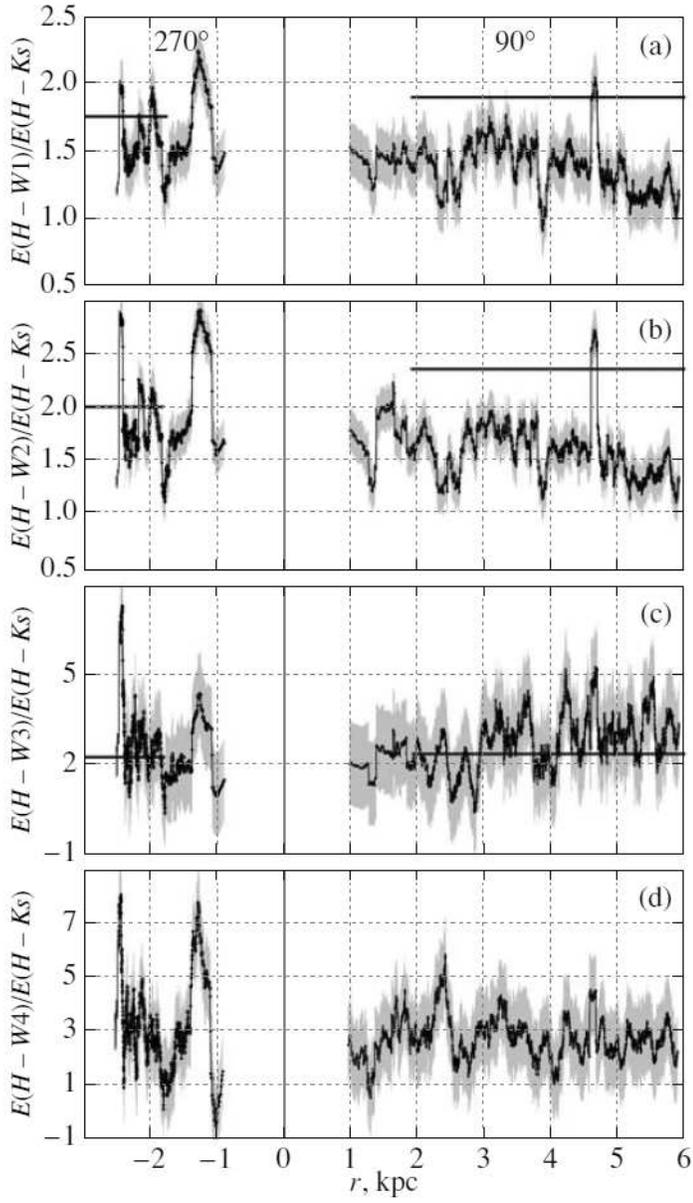}
\caption{Same as Fig. 3 for longitudes of $270^{\circ}$ (negative $r$)
and $90^{\circ}$ (positive $r$).
}
\label{xy}
\end{figure}

\begin{figure}
\includegraphics{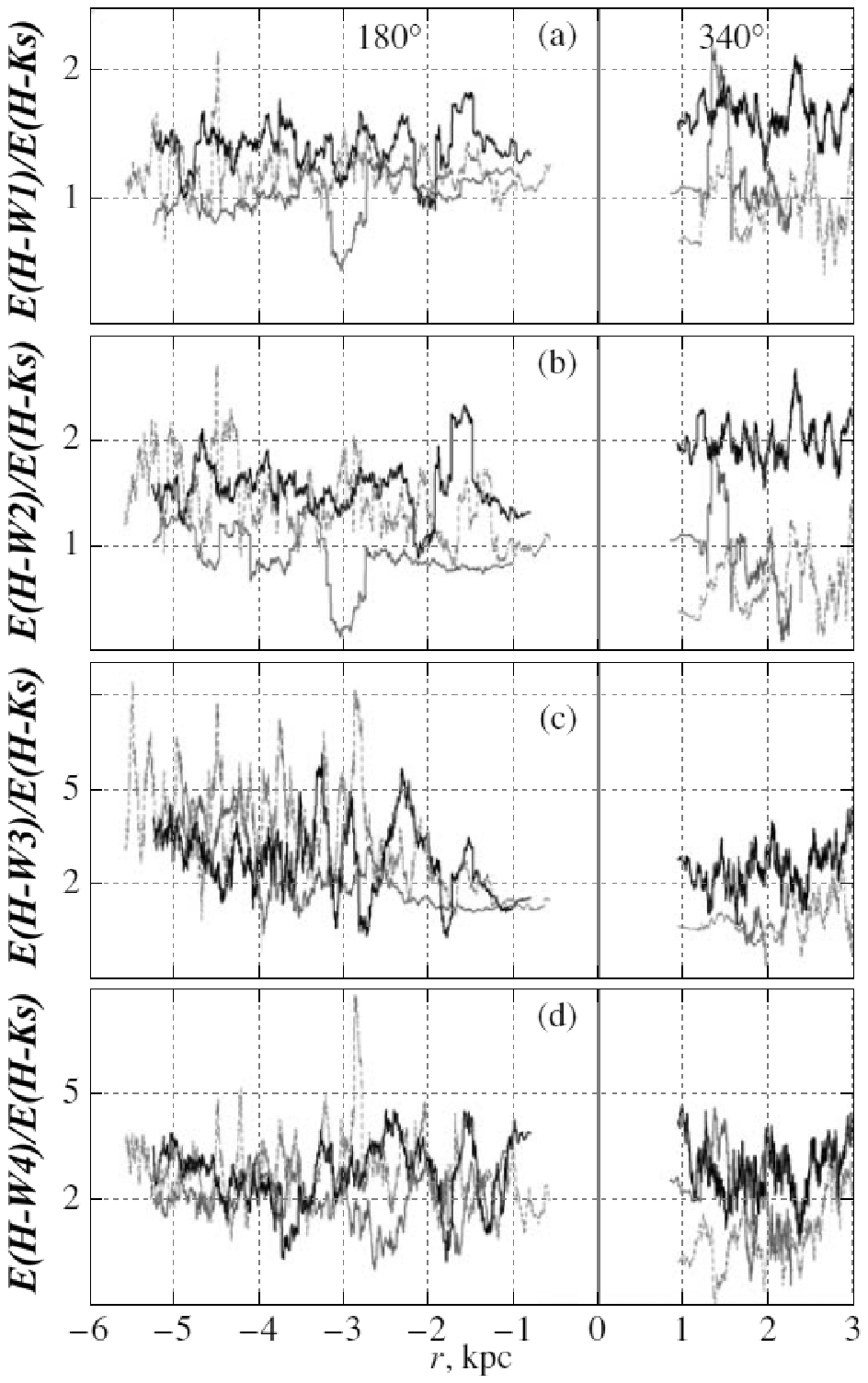}
\caption{Same as Fig. 3 for $l=180^{\circ}$, $b=0^{\circ}$ (black curves,
negative $r$), $l=180^{\circ}$, $b=+10^{\circ}$ (gray solid curves, negative
$r$), $l=180^{\circ}$, $b=-10^{\circ}$ (gray dashed curves, negative
$r$), $l=340^{\circ}$, $b=0^{\circ}$ (black curves, positive $r$), $l=340^{\circ}$, $b=+10^{\circ}$
(gray solid curves, positive $r$), $l=340^{\circ}$, $b=-10^{\circ}$ (gray dashed curves, positive $r$).
}
\label{lowhigh}
\end{figure}


\begin{thebibliography}{99}

\bibitem{benjamin} R.A.~Benjamin, E.~Churchwell, B.L.~Babler, et al., Publ. Astron. Soc. Pacif. \textbf{115}, 953 (2003).

\bibitem{bg2008} G.~Bertelli, L.~Girardi, P.~Marigo, et al., Astron. Astrophys. \textbf{484}, 815 (2008).

\bibitem{cardelli} J.A.~Cardelli, G.C.~Clayton and J.S.~Mathis, Astrophys. J. \textbf{345}, 245 (1989).

\bibitem{draine} B.T.~Draine, Ann. Rev. Astron. Astrophys. \textbf{41}, 241 (2003).

\bibitem{dcl} R.~Drimmel, A.~Cabrera-Lavers and M.~Lopez-Corredoira, Astron. Astrophys. \textbf{409}, 205 (2003).

\bibitem{hip} ESA, \emph{Hipparcos and Tycho catalogues} (ESA, 1997).

\bibitem{frisch} P.C.~Frisch, J.M.~Dorschner, J.~Geiss, et al., Astrophys. J. \textbf{525}, 492 (1999).

\bibitem{rcg} G.A. Gontcharov, Astron. Lett. \textbf{34}, 785 (2008).

\bibitem{model} G.A. Gontcharov, Astron. Lett. \textbf{35}, 638 (2009a).

\bibitem{gould} G.A. Gontcharov, Astron. Lett. \textbf{35}, 780 (2009b).

\bibitem{map} G.A. Gontcharov, Astron. Lett. \textbf{36}, 584 (2010).

\bibitem{rv} G.A. Gontcharov, Astron. Lett. \textbf{38}, 12 (2012a).

\bibitem{av} G.A. Gontcharov, Astron. Lett. \textbf{38}, 87 (2012b).

\bibitem{gro} M.A.T.~Groenewegen, Astron. Astrophys. \textbf{488}, 935 (2008).

\bibitem{tycho2} E.~H\o g, C.~Fabricius, V.V.~Makarov, et al., Astron. Astrophys. \textbf{355}, L27 (2000).

\bibitem{ind} R.~Indebetouw, J.S.~Mathis, B.L.~Babler, et al., Astrophys. J. \textbf{619}, 931 (2005).

\bibitem{jb} H.L.~Jonhson and J.~Borgman, Bull. Astron. Inst. Netherlands \textbf{17}, 115 (1963).

\bibitem{kruger} H.~Kr\"uger, E.~Gr\"un, M.~Landgraf, et al., Planet. Space Sci. \textbf{49}, 1303 (2001).

\bibitem{hip2} F.~van Leeuwen, Astron. Astrophys. \textbf{474}, 653 (2007).

\bibitem{lc} M.~Lopez-Corredoira, A.~Cabrera-Lavers, F.~Garzon, et al., Astron. Astrophys. \textbf{394}, 883 (2002).

\bibitem{marigo} P.~Marigo, L.~Girardi, A.~Bressan, et al., Astron. Astrophys. \textbf{482}, 883 (2008).

\bibitem{marshall} D.J.~Marshall, A.C.~Robin, C.Reyle, et al., Astron. Astrophys. \textbf{453}, 635 (2006).

\bibitem{nishiyama} S.~Nishiyama, M.~Tamura, H.Hatano, et al., Astrophys. J. \textbf{696}, 1407 (2009).

\bibitem{rc} W.~Reis and W.J.B.~Corradi, Astron. Astrophys. \textbf{486}, 471 (2008).

\bibitem{ppmxl} S.~Roeser, M.~Demleitner and E.~Schilbach, Astron. J. \textbf{139}, 2440 (2010).

\bibitem{skor} W.~Skorzynski, A.~Strobel and G.~Galazutdinov, Astron. Astrophys. \textbf{408}, 297 (2003).

\bibitem{2mass} M.F.~Skrutskie, R.M.~Cutri, R.~Stiening, et al.,
Astron. J. \textbf{131}, 1163 (2006); http://www.ipac.caltech.edu/2mass/releases/allsky/index.html.

\bibitem{dustmodel} J.C.~Weingartner and B.T.~Draine, Astrophys. J. \textbf{548}, 296 (2001).

\bibitem{wise} E.L.~Wright, P.R.M.~Eisenhardt, A.K.~Mainzer et. al., Astron. J. \textbf{140}, 1868 (2010),
http://irsa.ipac.caltech.edu/Missions/wise.html

\bibitem{zas} G.~Zasowski, S.R.~Majewski, R.~Indebetouw, et al., Astrophys. J. \textbf{707}, 510 (2009).


\end{thebibliography}
\end{document}